\documentclass[zpreprint,zbstepj]{zeus_paper}

\usepackage[english]{babel}

\chardef\usc=95
\chardef\til=126
\catcode`\@=11 
\DeclareRobustCommand\xdotspace{\futurelet\@let@token\@xdotspace}
\def\@xdotspace{%
  \ifx\@let@token.\else
  \ifx\@let@token\bgroup.\else
  \ifx\@let@token\egroup.\else
  \ifx\@let@token\/.\else
  \ifx\@let@token\ .\else
  \ifx\@let@token~.\else
  \ifx\@let@token!.\else
  \ifx\@let@token,.\else
  \ifx\@let@token:.\else
  \ifx\@let@token;.\else
  \ifx\@let@token?.\else
  \ifx\@let@token/.\else
  \ifx\@let@token'.\else
  \ifx\@let@token).\else
  \ifx\@let@token-.\else
  \ifx\@let@token\@xobeysp.\else
  \ifx\@let@token\space.\else
  \ifx\@let@token\@sptoken.\else
   .\space
   \fi\fi\fi\fi\fi\fi\fi\fi\fi\fi\fi\fi\fi\fi\fi\fi\fi\fi}
\catcode`\@=12 

\newcommand{\stru}[2]{%
   \relax\ifmmode\hbox{\vrule height#1 depth#2 width0pt}%
   \else\vrule height#1 depth#2 width0pt\fi}

\newcommand{\Ronum}[1]{\uppercase\expandafter{\romannumeral#1}}
\newcommand{\ronum}[1]{\expandafter{\romannumeral#1}}
\DeclareRobustCommand{\LaTeXZ}{%
  \LaTeX\kern-.05em4\kern-.1em
  {\raisebox{-0.2ex}{$\scriptstyle\text{ZEUS}$}}\xspace}

\DeclareMathAlphabet{\mathbf}{OT1}{cmr}{bx}{sl}
\newcommand{\eVdist}{\kern-0.06667em}

\newcommand{\Gev}{{\text{Ge}\eVdist\text{V\/}}}

\newcommand{\gev}{{\,\text{Ge}\eVdist\text{V\/}}}

\newcommand{\slashfrac}[2]{%
  \raisebox{0.5ex}{\ensuremath #1}\kern-0.12em/\kern-0.08em
  \raisebox{-.8ex}{\ensuremath #2}}

\newcommand{\sqr}[3]{%
    {\vcenter{\hrule height.#3ex\hbox{\vrule width.#2ex height#1ex
     \kern#1ex\vrule width.#3ex}\hrule height.#2ex}}}

\catcode`\@=11 
\newcommand{\parenbar}{\mathpalette\p@renb@r}
\def\p@renb@r#1#2{\vbox{%
  \ifx#1\scriptscriptstyle \dimen@.7em\dimen@ii.2em\else
  \ifx#1\scriptstyle \dimen@.8em\dimen@ii.25em\else
  \dimen@1em\dimen@ii.4em\fi\fi \offinterlineskip
  \ialign{\hfill##\hfill\cr
    \vbox{\hrule width\dimen@ii}\cr
    \noalign{\vskip-.3ex}%
    \hbox to\dimen@{$\mathchar300\hfil\mathchar301$}\cr
    \noalign{\vskip-.3ex}%
    $#1#2$\cr}}}
\catcode`\@=12 

\newcommand{\IP}{{\rm I$\kern-0.01667em$P}\xspace}

\mathchardef\qsm=63
\mathchardef\pls=43
\mathchardef\mns=512
\mathchardef\plm=518
\mathchardef\eql=61
\mathchardef\smallleft=300
\mathchardef\smallright=301
\mathchardef\les=316
\mathchardef\gre=318
\mathchardef\leq=532
\mathchardef\grq=533

\catcode`\@=11 
\newcounter{pict@width}
\newcounter{pict@height}
\newlength{\pict@scale}
\newcommand{\psfigadd}[4]{%
\setcounter{pict@width}{1*\ratio{#2+\pict@scale/2}{\pict@scale}}
\setcounter{pict@height}{1*\ratio{#3+\pict@scale/2}{\pict@scale}}
\setlength{\unitlength}{\pict@scale}
\hbox to #2{\hspace{-\fill}\begin{picture}(\thepict@width,\thepict@height)
\put(0,0){\psfig{figure=#1,width=#2,height=#3,clip=}}
\SetScale{0.283466457}
\SetWidth{1.763889}
{#4}
\end{picture}}
}
\newcounter{pict@widthfst}
\newcounter{pict@widthscd}
\newcounter{pict@widthtot}
\newcommand{\psfigaddtwo}[7]{%
\setcounter{pict@widthfst}{1*\ratio{#2+\pict@scale/2}{\pict@scale}}
\setcounter{pict@widthscd}{1*\ratio{#2+#4+\pict@scale/2}{\pict@scale}}
\setcounter{pict@widthtot}{1*\ratio{#2+#4+#6+\pict@scale/2}{\pict@scale}}
\setcounter{pict@height}{1*\ratio{#3+\pict@scale/2}{\pict@scale}}
\setlength{\unitlength}{\pict@scale}
\hbox{\hspace{-\fill}\begin{picture}(\thepict@widthtot,\thepict@height)
\put(0,0){\psfig{figure=#1,width=#2,height=#3,clip=}}
\put(\thepict@widthscd,0){\psfig{figure=#5,width=#6,height=#3,clip=}}
\SetScale{0.283466457}
\SetWidth{1.763889}
{#7}
\end{picture}}
}
\newcommand{\psfigror}[4]{%
\setcounter{pict@width}{1*\ratio{#2+\pict@scale/2}{\pict@scale}}
\setcounter{pict@height}{1*\ratio{#3+\pict@scale/2}{\pict@scale}}
\setlength{\unitlength}{\pict@scale}
\hbox{\begin{picture}(\thepict@width,\thepict@height)
\put(0,\thepict@height){\psfig{figure=#1,width=#3,height=#2,clip=,angle=270}}
\SetScale{0.283466457}
\SetWidth{1.763889}
{#4}
\end{picture}}
}
\newcommand{\psfigrol}[4]{%
\setcounter{pict@width}{1*\ratio{#2+\pict@scale/2}{\pict@scale}}
\setcounter{pict@height}{1*\ratio{#3+\pict@scale/2}{\pict@scale}}
\setlength{\unitlength}{\pict@scale}
\hbox{\begin{picture}(\thepict@width,\thepict@height)
\put(0,0){\psfig{figure=#1,width=#3,height=#2,clip=,angle=90}}
\SetScale{0.283466457}
\SetWidth{1.763889}
{#4}
\end{picture}}
}
\catcode`\@=12 
\newlength\listtextwidth

\catcode`\@=11 
\newlength{\@tabfninsert}
\newlength{\@tabfnwidth}
\newcommand{\tabfootnote}[2]{%
  \setlength{\@tabfninsert}{0.8em}
  \setlength{\@tabfnwidth}{\textwidth}
  \addtolength{\@tabfnwidth}{-\@tabfninsert}
  \addtolength{\@tabfnwidth}{-0.4em}
  \noindent\makebox[\@tabfninsert][r]{\footnotesize$^{#1}$\hfil}\hfill%
  \parbox[t]{\@tabfnwidth}{\footnotesize #2\hfill}}
\catcode`\@=12 

\def\citeCAL{{\cite{%
nim:a309:77,*nim:a309:101,*nim:a321:356,*nim:a336:23%
}}\xspace}

\begin{document}
\prepnum{{DESY--08--132}}
\def\lsim{\raisebox{-.65ex}{\rlap{$\sim$}} \raisebox{.4ex}{$<$}}
\def\gsim{\raisebox{-.65ex}{\rlap{$\sim$}} \raisebox{.4ex}{$>$}}

\title{
A measurement of the \boldmath{$Q^2$}, \boldmath{$W$} and \boldmath{$t$} 
dependences of deeply virtual Compton scattering at HERA
}                                                       
                    
\author{ZEUS Collaboration}
\date{December 2008}

\abstract{ 
Deeply virtual Compton scattering, $\gamma^* p \to \gamma p$, has been 
measured in $e^+p$ collisions at HERA with the ZEUS detector using an 
integrated luminosity of 61.1~pb$^{-1}$. 
Cross sections are presented as a function of the photon virtuality, 
$Q^2$, and photon-proton centre-of-mass energy, $W$, for a wide region of 
the phase space, $Q^2>$~1.5~GeV$^2$ and $40<W<170$ GeV. 
A subsample of events in which the scattered proton is measured in the leading 
proton spectrometer, corresponding to an integrated luminosity of 31.3 pb$^{-1}$,
is used for the first direct measurement of the differential cross section as a 
function of $t$, where $t$ is the square of the four-momentum transfer at the 
proton vertex.
}

\makezeustitle

\def\3{\ss}                                                                                        
\pagenumbering{Roman}                                                                              
                                                   %
\begin{center}                                                                                     
{                      \Large  The ZEUS Collaboration              }                               
\end{center}                                                                                       
  S.~Chekanov,                                                                                     
  M.~Derrick,                                                                                      
  S.~Magill,                                                                                       
  B.~Musgrave,                                                                                     
  D.~Nicholass$^{   1}$,                                                                           
  \mbox{J.~Repond},                                                                                
  R.~Yoshida\\                                                                                     
 {\it Argonne National Laboratory, Argonne, Illinois 60439-4815, USA}~$^{n}$                       
\par \filbreak                                                                                     
  M.C.K.~Mattingly \\                                                                              
 {\it Andrews University, Berrien Springs, Michigan 49104-0380, USA}                               
\par \filbreak                                                                                     
  P.~Antonioli,                                                                                    
  G.~Bari,                                                                                         
  L.~Bellagamba,                                                                                   
  D.~Boscherini,                                                                                   
  A.~Bruni,                                                                                        
  G.~Bruni,                                                                                        
  G.~Cara~Romeo                                                                                    
  F.~Cindolo,                                                                                      
  M.~Corradi,                                                                                      
\mbox{G.~Iacobucci},                                                                               
  A.~Margotti,                                                                                     
  T.~Massam,                                                                                       
  R.~Nania,                                                                                        
  A.~Polini\\                                                                                      
  {\it INFN Bologna, Bologna, Italy}~$^{e}$
% ...
\newpage                                                                                           
\enlargethispage{5cm}                                                                              
$^{\    1}$ also affiliated with University College London,                                        
United Kingdom\\                                                                                   
$^{\    2}$ now at University of Salerno, Italy \\                                                 
% ...
$^{\    5}$ supported by the research grant no. 1 P03B 04529 (2005-2008) \\                        
% ...
                    
\newpage   
\begin{tabular}[h]{rp{14cm}}                                                                                        
$^{a}$ &  supported by the Natural Sciences and Engineering Research Council of Canada (NSERC) \\  
$^{b}$ &  supported by the German Federal Ministry for Education and Research (BMBF), under        
          contract numbers 05 HZ6PDA, 05 HZ6GUA, 05 HZ6VFA and 05 HZ4KHA\\                         
$^{c}$ &  supported in part by the MINERVA Gesellschaft f\"ur Forschung GmbH, the Israel Science   
          Foundation (grant no. 293/02-11.2) and the U.S.-Israel Binational Science Foundation \\  
$^{d}$ &  supported by the Israel Science Foundation\\                                             
$^{e}$ &  supported by the Italian National Institute for Nuclear Physics (INFN) \\                
$^{f}$ &  supported by the Japanese Ministry of Education, Culture, Sports, Science and Technology 
          (MEXT) and its grants for Scientific Research\\                                          
$^{g}$ &  supported by the Korean Ministry of Education and Korea Science and Engineering          
          Foundation\\                                                                             
$^{h}$ &  supported by the Netherlands Foundation for Research on Matter (FOM)\\                   
$^{i}$ &  supported by the Polish State Committee for Scientific Research, project no.             
          DESY/256/2006 - 154/DES/2006/03\\                                                        
$^{j}$ &  partially supported by the German Federal Ministry for Education and Research (BMBF)\\   
$^{k}$ &  supported by RF Presidential grant N 1456.2008.2 for the leading                         
          scientific schools and by the Russian Ministry of Education and Science through its      
          grant for Scientific Research on High Energy Physics\\                                   
$^{l}$ &  supported by the Spanish Ministry of Education and Science through funds provided by     
          CICYT\\                                                                                  
$^{m}$ &  supported by the Science and Technology Facilities Council, UK\\                         
$^{n}$ &  supported by the US Department of Energy\\                                               
$^{o}$ &  supported by the US National Science Foundation. Any opinion,                            
findings and conclusions or recommendations expressed in this material                             
are those of the authors and do not necessarily reflect the views of the                           
National Science Foundation.\\                                                                     
$^{p}$ &  supported by the Polish Ministry of Science and Higher Education                         
as a scientific project (2006-2008)\\                                                              
$^{q}$ &  supported by FNRS and its associated funds (IISN and FRIA) and by an Inter-University    
          Attraction Poles Programme subsidised by the Belgian Federal Science Policy Office\\     
$^{r}$ &  supported by an FRGS grant from the Malaysian government\\       
\end{tabular}                                                  

\newpage

\pagenumbering{arabic} 
\pagestyle{plain}

\section{Introduction}
\label{sec-int}
 
This paper presents cross-section measurements for the exclusive
production of a real photon in diffractive $e^+p$\footnote{Hereafter, 
the positron is referred to with the same symbol, $e$, as the electron.} 
interactions,
$ep \rightarrow e\gamma p$, a process known as deeply virtual Compton
scattering (DVCS). In perturbative QCD, this process is described by
the exchange of two partons, with different longitudinal and transverse
momenta in a colourless configuration. At the $\gamma ^* p$ centre-of-mass 
energies,
$W$, available for $ep$ collisions at the HERA collider, for large
momentum-transfer squared at the lepton vertex, $Q^2$, the DVCS
process is dominated by two-gluon exchange. Measurements of the DVCS
cross section provide constraints on the generalised parton
distributions
(GPDs)~\cite{Diehl:1997bu,Burkardt:2003,Diehl:2002,Guzey:2006,Kumericki:2008},
which carry information about the wave function of the proton)~\cite{Burkardt:2003}. The
transverse distribution of partons in the proton, which is not
accessible via the $F_2$ proton structure function,
is accounted for in the dependence of the GPDs on the four-momentum transfer
squared at the proton vertex, $t$.  The initial and
final states of the DVCS process are identical to those of the purely
electromagnetic Bethe-Heitler (BH) process.  The interference between
these two processes in principle provides information about the real
and imaginary parts of the QCD scattering
amplitude~\cite{plb:460:417:1999,Belitsky:2001ns,Freund:2001hm}.
However, the interference is expected to be small in the kinematic 
region studied in this paper~\cite{plb:460:417:1999,Belitsky:2001ns}.

The simplicity of the final state and the absence of complications due
to hadronisation mean that the QCD predictions for DVCS are expected to be more
reliable than for exclusive vector meson 
production which has been extensively studied in $ep$ collisions at
HERA~\cite{pl:b487:273,np:b718:3,np:b695:3,epj:c14:213,pl:b483:360,art:pmc:a2007:1,art:epj:c46:585,epj:c13:371}. 
Several measurements of DVCS  at high
$W$ are available~\cite{pl:b517:47,pl:b573:46,art:epj:c44:1,art:pl:c659:796}. 
The analysis presented here is based on data in the kinematic range of 
$1.5<Q^2<100\,\Gev^2$ and $40<W<170 \,\Gev$, an extension compared to the
previous ZEUS measurement~\cite{pl:b573:46}.
A subsample of the data in which
the scattered proton is measured in the ZEUS leading proton
spectrometer (LPS)~\cite{zfp:c73:253} is used for the direct
measurement of the $t$ dependence of the DVCS cross section.

\section{Experimental set-up}

The data used for this measurement were taken with the ZEUS detector at
the HERA $ep$ collider in the years 1999 and 2000, when HERA collided
positrons of energy 27.5 GeV with protons of energy 920~GeV, and correspond 
to an integrated luminosity of 61.1 pb$^{-1}$.
The subsample used to measure the $t$ distribution was collected in 2000 and 
corresponds to an integrated luminosity of  31.3 pb$^{-1}$.  

A detailed description of the ZEUS detector can be found
elsewhere~\cite{zeus:1993:bluebook,pl:b293:465}. A brief outline of the 
components most relevant for this analysis is given below.

Charged particles were tracked in the
CTD~\cite{nim:a279:290,*npps:b32:181,*nim:a338:254}. The CTD
operated in a magnetic field of 1.43 T provided by a thin solenoid.
It consisted of
72 cylindrical drift-chamber layers, organised in nine superlayers
covering the polar-angle~\footnote{The ZEUS coordinate system is a
right-handed Cartesian system, with the $Z$ axis pointing in the proton
direction, referred to as the ``forward direction'', and the $X$ axis
pointing left towards the centre of HERA. The coordinate origin is at the
nominal interaction point. The pseudorapidity is defined as 
\mbox{$\eta =-\ln{(\tan{\frac{\theta}{2}})}$}, where the polar angle 
$\theta$ is measured with respect to the proton beam direction.}
region $15^{\circ}<\theta<164^{\circ}$. 
The transverse-momentum resolution for full-length tracks
was $\sigma(p_T)/p_T = 0.0058p_T \oplus 0.0065 \oplus 0.0014/p_T$, with
$p_T$ in GeV.

The uranium--scintillator calorimeter (CAL)~\citeCAL covered 99.7\% of the 
total solid angle and consisted of three 
parts: the forward (FCAL), the barrel
(BCAL) and the rear (RCAL) calorimeters. Each part was subdivided
transversely into towers and longitudinally into one electromagnetic
section (EMC) and either one (in RCAL) or two (in BCAL and FCAL) hadronic
sections (HAC). The CAL energy resolutions, as measured under 
test-beam conditions, were $\sigma(E)/E=0.18/\sqrt{E}$ for positrons 
and $\sigma(E)/E=0.35/\sqrt{E}$ for hadrons, with $E$ in $\Gev$. 

The position of positrons scattered at small angles to the
positron-beam direction was determined combining the information from
the CAL, the small-angle rear tracking detector (SRTD) 
and the hadron-electron separator (HES)~\cite{nim:a401:63,nim:a277:176}. 

The FPC~\cite{nim:a450:235} was used to measure the energy of
particles in the pseudorapidity range $\eta \approx 4.0 - 5.0$. It
consisted of a lead--scintillator sandwich calorimeter installed in the
$20\times 20$ cm$^2$ beam hole of the FCAL.  The energy resolution for
electrons as measured in a test beam, was $\sigma(E)/E = (0.41 \pm
0.02)/\sqrt{E} \oplus 0.062\pm 0.002$,  with $E$ in GeV.   
The energy resolution for pions was $\sigma(E)/E = (0.65 \pm
0.02)/\sqrt{E} \oplus 0.06 \pm 0.01$, with $E$ in GeV, 
after having combined the information from FPC and FCAL. 
The $e/h$ ratio was close to unity.

The LPS~\cite{zfp:c73:253} detected positively charged particles scattered at 
small angles and carrying a substantial fraction, $x_L$, of the incoming
proton momentum; these particles remained in the beam-pipe and their
trajectory was measured by a system of silicon microstrip detectors
that was inserted very close (typically within a distance of a few mm)
to the proton beam.  The detectors were grouped in six stations, S1 to
S6, placed along the beam-line in the direction of the proton beam,
between 23.8 m and 90.0 m from the interaction point. The particle
deflections induced by the magnets of the proton beam-line allowed a
momentum analysis of the scattered proton. For the present
measurements, only stations S4, S5 and S6 were used. The resolutions
were about $0.5\%$ on the longitudinal momentum fraction and about 5
MeV on the transverse momentum. The LPS 
acceptance~\cite{art:epj:c38:43} was approximately
2\% and $x_L$-independent for $x_L \gsim 0.98$; it increased smoothly
to about 10\% as $x_L$ decreased to 0.9.

The luminosity was measured from the rate of the bremsstrahlung process 
$ep \rightarrow e\gamma p$, where the photon was measured in a 
lead--scintillator calorimeter~\cite{desy-92-066,*zfp:c63:391,*acpp:b32:2025}
placed in the HERA tunnel at $Z=-107$~m.

\section{Monte Carlo simulations}

The acceptance and the detector response were determined using Monte
Carlo (MC) simulations. The detector was simulated in detail using a
program based on {\sc Geant~3.13}~\cite{misc:cern:geant3}. All of the
simulated events were processed through the same reconstruction and
analysis chain as the data.

A MC generator, {\sc GenDVCS}~\cite{misc:gendvcs}, based on a model by
Frankfurt, Freund and Strikman (FFS)~\cite{pr:d58:114001}, was used to
simulate the elastic DVCS process as described in~\cite{pl:b573:46}.  
The ALLM97~\cite{hep-ph-9712415} parameterisation of the
$F_2$ proton structure function of the proton was used as input. 
The $t$ dependence was assumed to be exponential with a
slope parameter $b$ set to 4.5~$\rm{GeV^{-2}}$, independent of $W$ and $Q^2$.

The elastic, $ep\rightarrow e\gamma p$, and quasi-elastic 
$ep\rightarrow e\gamma Y$ BH processes,
where $Y$ is a low-mass state, and the exclusive dilepton production,
$ep\rightarrow ee^+e^-p$, were simulated using the
{\sc Grape-Compton}\footnote{Hereafter, the {\sc Grape-Compton} generator is
referred to as {\sc Grape}.}~\cite{cpc:136:126} and the
{\sc Grape-Dilepton}~\cite{cpc:136:126} generators. These two
MC programs are based on the automatic system
{\sc Grace}~\cite{misc:kek:grape-compton} for calculating Feynman diagrams.
A possible contribution from vector meson electroproduction was simulated with
the {\sc Zeusvm} generator~\cite{thesis:muchorowski:1998}.
To account for electroweak radiative effects, all the generators were
interfaced to {\sc Heracles~4.6}~\cite{spi:www:heracles}. 

\section{Kinematic variables and event selection}
\label{sec-strategy}

The process $ep\rightarrow e\gamma p$ is parametrised by the following 
variables: 

\begin{itemize}

\item $Q^2=-q^2=-(k-k')^2$, the negative 
four-momentum squared of the virtual photon, where $k~(k')$ is the 
four-momentum of the incident (scattered) positron;
\item $W^2=(q+p)^2$, the squared centre-of-mass energy of the photon-proton 
system, where $p$ is the four-momentum of the incident proton;
\item $x=Q^2/(2P\cdot q)$, the fraction of the proton momentum carried 
by the quark struck by the virtual photon in the infinite-momentum 
frame (the Bjorken variable);
\item $x_L=\frac {p' \cdot k}{p \cdot k}$, the fractional momentum of the 
outgoing proton, where $p'$ is the four-momentum of the scattered proton;
\item $t=(p-p')^2$, the squared four-momentum transfer at the proton vertex.

\end{itemize}

For the $Q^2$ range of this analysis, $Q^2> 1.5$ GeV$^2$, and at small
values of $t$, the signature of elastic DVCS and BH events consists of a 
scattered positron, a photon and a scattered proton.  
The scattered proton remains in the beam-pipe where, for a subsample of events,
it is detected in the LPS (LPS sample).

The events were selected online via a three-level trigger
system~\cite{zeus:1993:bluebook,Smith:1992}. The trigger required
events with two isolated electromagnetic (EM) clusters with energy
greater than 2 \Gev. The trigger efficiency was studied as a function of the 
lowest energy cluster, it was found to increase from 80\% to 100\% for 
increasing cluster energy and the Monte Carlo was reweighed according.

The offline selection followed the strategy described in~\cite{pl:b573:46}. 
Two EM clusters were found by a dedicated, neural-network based, positron
finder~\cite{nim:a365:508,*nim:a391:360}. 
They were ordered in polar-angle and are in the following denoted 
as EM1 and EM2, with $\theta_1>\theta_2$.
The first cluster was
required to be in the RCAL with energy $E_1>10 \,\Gev$; the second cluster
had to have a polar angle $\theta_2 <2.85
\,\mathrm{rad}$ and  was required to
be either in the RCAL, with energy $E_2>3 \,\Gev$, or in the BCAL,
with energy $E_2>2.5 \,\Gev$. The angular range of the second cluster
corresponds to the region of high reconstruction efficiency for tracks
in the CTD. The association of a track discriminates between positron and 
photon induced clusters.
For events with one
track, a match was required between the track and one of the two EM
clusters. Events with more than one track were rejected.
To ensure full containement of the electromagnetic shower, the impact 
position of each EM cluster on the face of RCAL was required to be outside a 
rectangular area of $26\times 16$~cm$^2$ around the beam-pipe.
 
The condition $40<E-P_Z<70\;\Gev$ was imposed, with $E=E_1+E_2$ and
$P_Z=E_1 \cos \theta_1 + E_2 \cos \theta_2 $.  
This requirement rejected photoproduction events and also
events in which a hard photon was radiated from the incoming positron.

Events with CAL energy deposits not associated with the two EM clusters were 
rejected if their energy was above the noise level in the 
CAL~\cite{thesis:ciesielski:2004}.  
In addition, the total energies measured in the FPC and in
the FCAL were each required to be below 1~GeV~\cite{np:b718:3,thesis:ciesielski:2004}.  
These elasticity requirements also suppressed DVCS events and inelastic 
BH events in which the proton
dissociates into a high-mass hadronic system.
The sample was still contaminated by events in which a forward,  low-mass hadronic 
system was not visible in the main detector.
Alternatively, a clean sample of elastic DVCS and BH
events was obtained by additionally requiring the proton to be
detected in the LPS. 

The LPS event was rejected if, at any point, the distance of the proton track candidate
to the beam-pipe was less than 0.04~cm. 
It was also rejected if the $X$ position of the track impact point at
station S4 was smaller than $-3.3$~cm. These cuts reduced the
sensitivity of the acceptance on the uncertainty in the position of
the beam-pipe apertures. 
To suppress background from overlays of $ep$ collisions with
protons originating from the beam-halo, it was required that 
$(E+P_Z) +2 p_Z^{\rm LPS} <1865$~GeV, where $p_Z^{\rm LPS}$ is the 
longitudinal momentum of the scattered proton. 
The variable $x_L$ was required to be within the range
$0.96<x_L<1.02$ to exclude non-elastic events~\cite{desy-08-176,pmc:a1:6}. 
The variable $t$ was required to be in the range $0.08<|t|<0.53$~GeV$^2$
where the LPS acceptance was high and slowly changing.  

The kinematic region was 
\mbox{$40 <W<170 \,\Gev$} and \mbox{$1.5 < Q^2 < 100\,\Gev^2$}.
For the purposes of this analysis, the values of $Q^2$ and $W$ were
determined for each event, independently of its topology, under the
assumption that the EM1 cluster is the scattered positron. 
This assumption is always valid for DVCS events for the $Q^2$ range 
considered here.
The electron method~\cite{proc:hera:1991:23,*hoeger} was used to 
determine $Q^2$ and the double-angle 
method~\cite{proc:hera:1991:23,*hoeger} to determine $W$.

\section{Background study and signal extraction}
\label{sec-analysis}

The selected events were subdivided into three samples, 

\begin{itemize}
\item $\gamma$ sample: EM2, with no track pointing to it, is taken to be the
  photon and EM1 is assumed to be the scattered positron. Both  
  BH and DVCS processes contribute to this topology. The sample  
  consisted of 7618 events and 55 events after the LPS selection.

\item $e$ sample: EM2, with a positive-charge track pointing to it, is
  assumed to be the scattered positron and EM1 is the photon.  
  The sample is dominated by BH events. The number of DVCS events is 
  predicted to be negligible due to the large $Q^2$ implied by the
  large positron scattering angle. 
  This sample consisted of 11988 events and 33 events after the LPS selection.
    
\item \mbox{negative-charge-$e$ sample}: EM2, with a negative-charge
  track pointing to it, may have originated from an $e^+ e^-$ final
  state accompanying the scattered positron, where one of the
  positrons escaped detection. This sample is dominated by
  non-resonant $e^+ e^-$ production and by $J/\psi$ production with
  subsequent decay into $e^+ e^-$ and was used to study these background 
  sources. It consisted of 764 events and only one event after the LPS
  selection.The diffractive electroproduction of $\rho$, $\omega$ and
  $\phi$ mesons was found to be negligible~\cite{pl:b573:46}. 

\end{itemize}

In the kinematic region of this analysis, the contribution of the
interference term between the DVCS and BH amplitudes is very small
when the cross section is integrated over the angle between the positron
and proton scattering planes~\cite{plb:460:417:1999,Belitsky:2001ns}.
Thus the cross section for exclusive production of real photons was
treated as a simple sum over the contributions from the DVCS and
BH processes. The DVCS cross section was determined by subtracting the latter.

The size of the BH contribution to be subtracted was determined using 
the $e$ sample
which consists of elastic and inelastic BH events and a small fraction
of exclusive $e^+e^-$ production.
The exclusive  $e^+e^-$ contribution was estimated with the 
negative-charge-$e$ 
sample to be $(6.4\pm 0.2)$\% and subtracted from the $e$ sample. 

The inelastic fraction  of the BH events was estimated from the difference in 
the azimuthal angles, $\Delta\phi$, between the two electromagnetic clusters 
in the $e$ sample. It was determined to be $(16 \pm 1)\%$ and was
negligible in the LPS tagged subsample~\cite{thesis:fazio:2007}. 

The measured cross section of the BH process was $(4\pm 1)$\%
smaller than the expectations of the {\sc Grape} program 
(a detailed discussion can be found 
elsewhere~\cite{thesis:bold:2003,thesis:fazio:2007}). 
The {\sc Grape} cross section was modified accordingly.

The BH contribution to the $\gamma$ sample was determined by {\sc Grape} and 
found to be $(56 \pm 1)$\% for the untagged and ($21\pm 3$)\% for the LPS 
tagged sample. The BH-subtracted $\gamma$ sample was further scaled by 
$(1-f_{p\rm{-diss}})$, where $f_{p\rm{-diss}}$ is the fraction of DVCS events 
in which the proton dissociated into a low-mass state. Its value was taken, 
as in~\cite{pl:b573:46}, from previous publication~\cite{epj:c24:345}, 
$f_{p\rm{-diss}}=17.5\pm1.3^{+3.7}_{-3.2}$\%.

\begin{figure}
\centering
\includegraphics[scale=0.75]{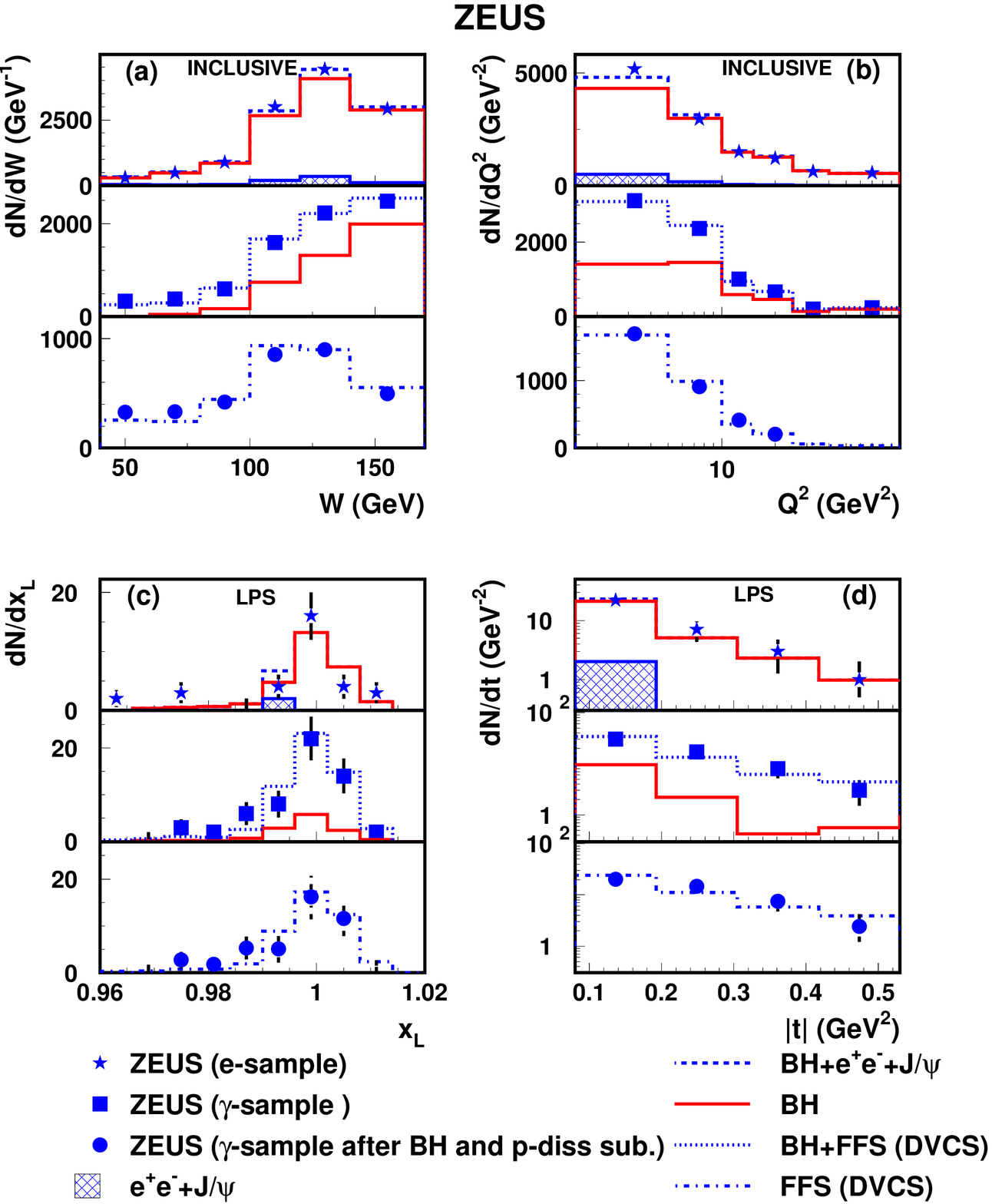}
\caption{
Distribution of (a) $W$, (b) $Q^2$ in the inclusive sample and of (c) $x_L$ and (d) $|t|$ in 
the LPS sample, for the $e$-sample (top), the $\gamma$-sample (middle) and the 
$\gamma$-sample after BH background and proton dissociation subtraction (bottom). 
Also shown are the expectations of the MC normalised to the luminosity of the data and the 
contribution from exclusive dilepton production ($e^+ e^-$).
}
\label{fig:dmc}
\end{figure}

The $W$ and $Q^2$ distributions in the untagged sample (inclusive sample) 
and the
$x_L$ and $t$ distributions in the LPS sample, separately for the $e$
sample, for the $\gamma$ sample and for the $\gamma$ sample after 
BH and proton dissociation background subtraction, 
are shown in Fig.~\ref{fig:dmc}. Also shown in
the figure are MC expectations which describe the data well.

\section{Systematic uncertainties}
\label{sec-syst}

The uncertainties due to the reconstruction of the scattered 
positron and to the background subtraction were evaluated by varying the
selection criteria as follows: 

\begin{itemize}
\item varying the electromagnetic energy scale by $\pm$2\%;
\item restricting the $E-P_Z$ cut to  $45<E-P_Z<65$~GeV;
\item shifting the reconstructed position of the positron with respect 
to the MC by $\pm1$mm; 
\item changing the elasticity requirements by $\pm30$~MeV in the EMC and
      $\pm 50$~MeV in the HAC sections;
\item changing the photon candidate energy by $\pm$10\%;
\item varying the inelastic BH fraction by $\pm$1\%.
\end{itemize}

Each individual systematic uncertainty affects single bins in $Q^2$ and $W$  
typically by less than 5\% and by less than 10\%  in all cases bar the
highest $Q^2$ bin where statistical fluctuations dominate.

To evaluate the uncertainties due to the reconstruction of the final-state 
proton,
\begin{itemize}
\item 
the cut on the minimum distance to the beam-pipe was increased to 0.1 cm; 
\item
the $t$ range was tightened to $0.1<|t|<0.5$~GeV$^2$;
\item
the kinematic limit of the beam-halo background cut was lowered to $1855$~GeV;
\item
the $x$ position of the track impact point at station S4 was restricted to 
$-32$~mm.
\end{itemize}

The total systematic uncertainty was obtained 
by adding in quadrature the individual contributions.
It was found to be $\pm$8\% on average which is smaller than
the statistical uncertainties. 
    
For the inclusive sample the uncertainty in the determination of the integrated 
luminosity of $\pm$2.25\% and on the proton-dissociative background of 
$^{+3.9}_{-3.5}\%$ are not included in the figures and in the tables.

For the LPS data, there is an overall uncertainty of $\pm 7\%$ which originates 
mostly from the uncertainty on the simulation of the proton-beam optics. 
It can be treated as a normalisation uncertainty as it is 
largely independent of the kinematic variables and is not included in 
the figures and in the tables. It also includes the uncertainty on the
integrated luminosity for the LPS sample of $\pm$2.25\%.

\section{Cross section determination and results}
\label{sec-xsec}

The $\gamma^\ast p$ cross section of the DVCS process was evaluated 
as a function of $W$, $Q^2$ using the expression 

\begin{displaymath}
\sigma^{\gamma^\ast p \rightarrow \gamma p}(W_i, Q^2_i)
=\frac{(N_i^{\rm{obs}}-N_i^{\rm{BH}})\cdot(1-f_{p\rm{-diss}})}
{N_i^{\rm{MC}}}\cdot 
\sigma ^{{\rm{FFS}} (\gamma^\ast p \rightarrow \gamma p)} (W_i, Q^2_i),    
\label{eq-ffs}   
\end{displaymath}
where $N^{\rm{obs}}_i$ is the total number of data events in the
$\gamma$ sample in bin $i$ of $W$ and $Q^2$, $N^{\rm{BH}}_i$ denotes
the number of elastic and inelastic BH events in the $\gamma$ sample in 
the bin, and
$N^{\rm{MC}}_i$ is the number of events expected in the $\gamma$
sample from {\sc GenDVCS} for the luminosity of the data. 
The cross section as predicted by the FFS model is denoted
$\sigma^{\rm{FFS}}(\gamma^\ast p \rightarrow \gamma p)$ and was
evaluated at the centre $(W_i, Q^2_i)$ of each $Q^2$ and $W$ bin.
The differential cross section as a function of $t$ was calculated 
from the LPS tagged sample for which $f_{p\rm{-diss}}$ is zero.
All the results are listed in Tables~\ref{tab:tab1}\,--\,\ref{tab:tab4}.

\begin{figure}[p]
\begin{center}
\includegraphics[width=15cm,height=15cm]{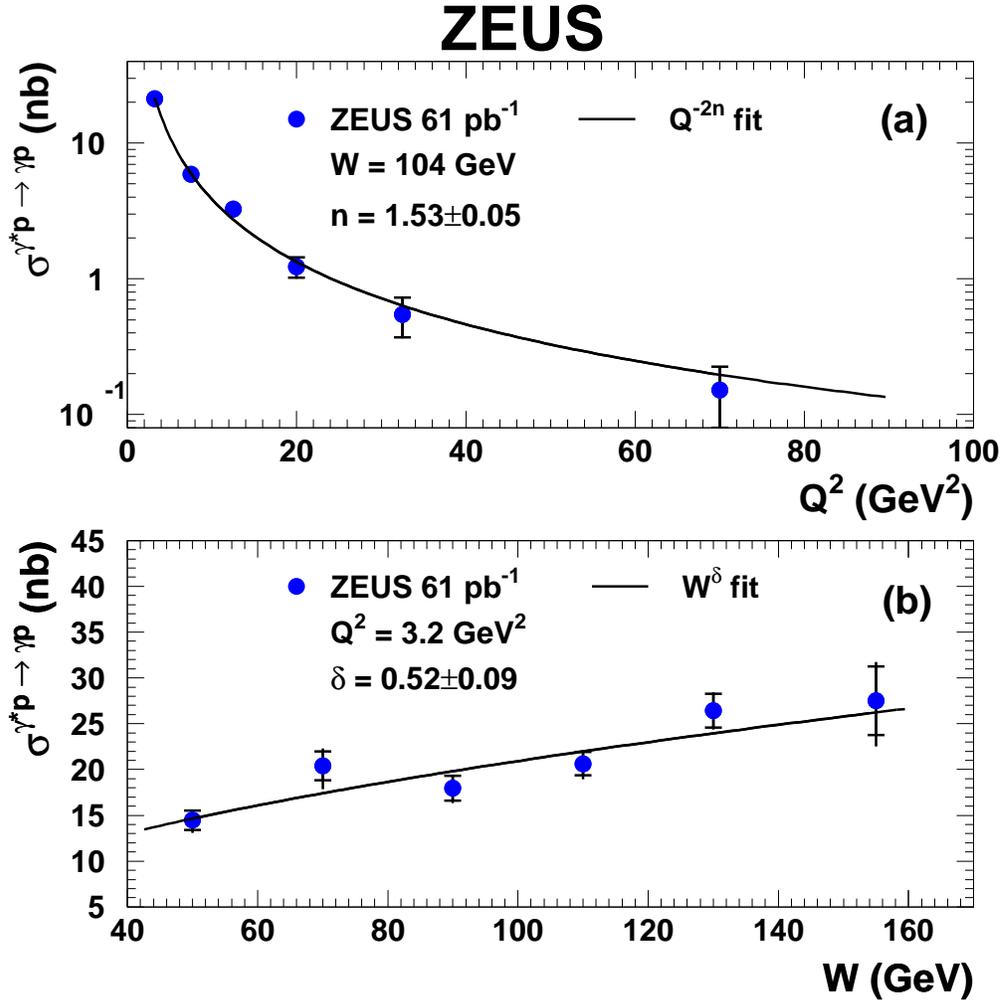}
\end{center}
\caption{
(a) The DVCS cross section, $\sigma^{\gamma ^*p\rightarrow \gamma p}$, 
as a function of $Q^2$.  
The solid line is the result of a fit of the form $\sim Q^{-2n}$.
(b) The DVCS cross section, $\sigma^{\gamma ^*p\rightarrow \gamma
p}$, as a function of $W$.
The solid line is the result of a fit of the form $\sim W^\delta$.
The inner error bars represent the statistical uncertainty while the outer
error bars the statistical and systematic uncertainties added in quadrature. 
}
\label{fig:q2andw}
\end{figure}

\begin{figure}
\centering
\includegraphics[width=15cm,height=15cm]{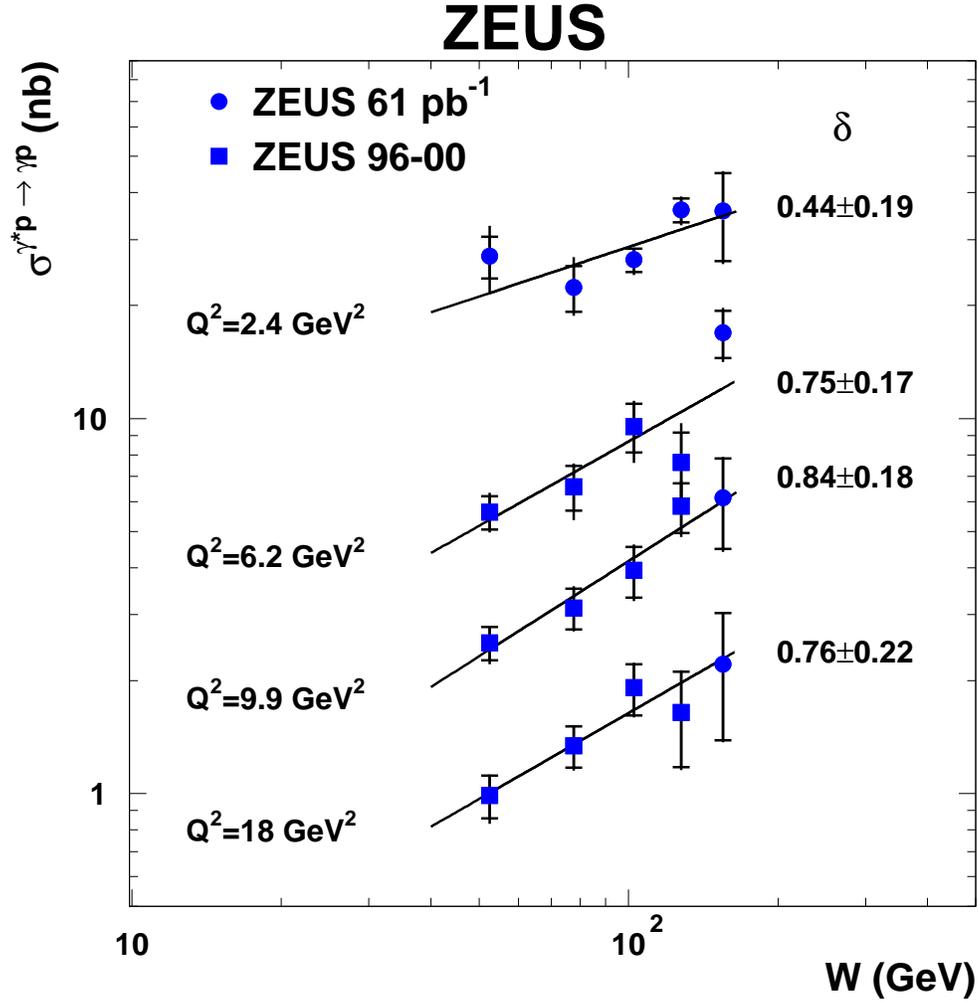}
\caption{
The DVCS cross section, $\sigma^{\gamma ^*p\rightarrow \gamma p}$, as a function of 
$~W$ for  $Q^2=2.4\;\Gev^2$ (dots) shown together with previous ZEUS measurements 
(squares)~\protect\cite{pl:b573:46}. 
Also shown at higher $Q^2$ are the new measurements at $W=155$~GeV (dots).
The solid lines are the results of a fit of the
form $\sigma^{\gamma ^*p\rightarrow \gamma p} \propto W^\delta$. 
The values of $\delta$ and their statistical uncertainties are given in the figure.  
The inner error bars represent the statistical uncertainty while the outer
error bars the statistical and systematic uncertainties added in quadrature. 
}
\label{fig:wfit}
\end{figure}

The $\gamma^\ast p$ DVCS cross section, 
$\sigma^{\gamma^\ast p \rightarrow \gamma p}$, 
is presented in Fig.~\ref{fig:q2andw} as a function of $Q^2$ at $W=104$~GeV 
and as a function of $W$ at $Q^2=3.2$~GeV${}^2$. 
The cross section
shows a fast decrease with $Q^2$.  A fit to the $Q^2$ dependence of
the cross section, assuming the functional form 
$\sigma^{\gamma^\ast p \rightarrow \gamma p}(Q^2)\sim Q^{-2n}$, was
performed for $W=104$~GeV yielding $n = 1.54 \pm 0.05 (\rm stat.)$, smaller 
than expected for a pure propagator term~\cite{pr:d58:114001}.
The result is in agreement with other DVCS measurements at
HERA at lower $W$~\cite{pl:b517:47,pl:b573:46,art:epj:c44:1,art:pl:c659:796}. 
As expected for DVCS~\cite{pr:d58:114001}, the decrease of the cross section
with $Q^2$ is slower than for exclusive vector meson
production~\cite{epj:c6:603,art:epj:c46:585,epj:c13:371,epj:c14:213,pl:b483:360,art:pmc:a2007:1}.
The cross section increases with $W$. In pQCD-based models, this
behaviour is related to the increase of the gluon content of the
proton with decreasing Bjorken-$x$.  A fit to the $W$ dependence of
the cross section, assuming a functional form 
$\sigma^{\gamma^\ast p \rightarrow \gamma p}(W) \sim W^\delta$, 
was performed for $Q^2=3.2$~GeV$^2$, yielding $\delta =0.52\pm 0.09 (\rm stat.)$. 
This result is in agreement with the previous 
measurements~\cite{pl:b517:47,pl:b573:46,art:epj:c44:1,art:pl:c659:796} 
performed in a restricted range of $W$ and at higher $Q^2$.
A second fit restricted to the region $1.5 < Q^2 < 5$~GeV$^2$ at
$Q^2=2.4$ GeV$^2$, was also performed giving 
$\delta =0.44\pm 0.19 (\rm stat.)$.  The fit is presented in
Fig.~\ref{fig:wfit}. Also shown in the figure are previous ZEUS
measurements at different values of $Q^2$~\cite{pl:b573:46} and 
the extension to higher $W$ values from the present analysis. 
For each $Q^2$ the corresponding $\delta$ values fitted in the extended $W$ 
range are given. 
Within the present accuracy the results do not show evidence for a $Q^2$
dependence of $\delta$. This result is similar to that obtained for
the exclusive production of $J/\psi$
mesons~\cite{epj:c24:345,epj:c14:213,pl:b483:360,art:pmc:a2007:1}.

\begin{figure}
\begin{center}
\includegraphics[scale=0.8]{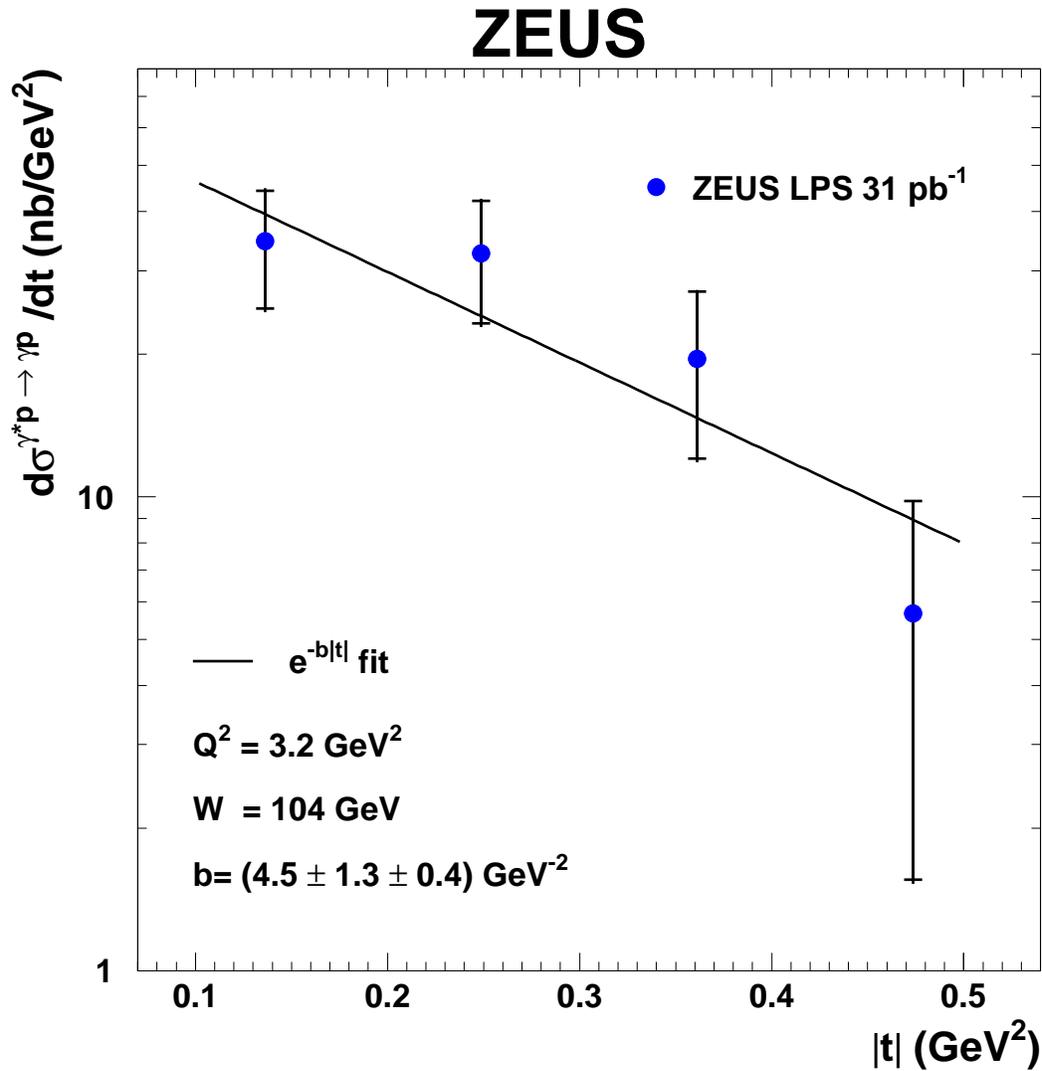}
\end{center}
\vspace{-1.2cm}
\caption{
The DVCS differential cross section, $d\sigma^{\gamma ^*p\rightarrow \gamma p}/dt$, 
as a function of $|t|$. 
The solid line is the result of a fit of the form $\sim e^{-b|t|}$.
The inner error bars represent the statistical uncertainty while the outer
error bars the statistical and systematic uncertainties added in quadrature. 
}
\label{fig:tfignew}
\end{figure}

The first direct measurement of the differential cross section
$d\sigma^{\gamma^\ast p\rightarrow \gamma p}/dt$, 
extracted from the LPS-tagged events at $Q^2=3.2$~GeV$^2$ and 
at $W=104$~GeV, is shown in Fig.~\ref{fig:tfignew}.
The value of the slope parameter $b$ extracted from an exponential fit 
to the differential cross section, 
$d\sigma ^{\gamma^\ast p \to \gamma p}/dt \propto  e^{-b|t|}$, is 
$b=4.5\pm 1.3 (\rm stat.) \pm 0.4 (\rm syst.)~ \rm{GeV^{-2}}$
($\chi^2/\mathrm{ndf}=0.90$).  
This value is consistent with the results obtained by 
H1~\cite{art:pl:c659:796} $b=5.45\pm 0.19 (\rm stat.) \pm 0.34
(\rm syst.)~\rm{GeV^{-2}}$ at $Q^2=8$~GeV$^2$ and $W=82$~GeV,
from the transverse-momentum distribution of the photon candidate.

\begin{figure}
\vspace{-.5cm}
\begin{center}
\includegraphics[scale=0.8]{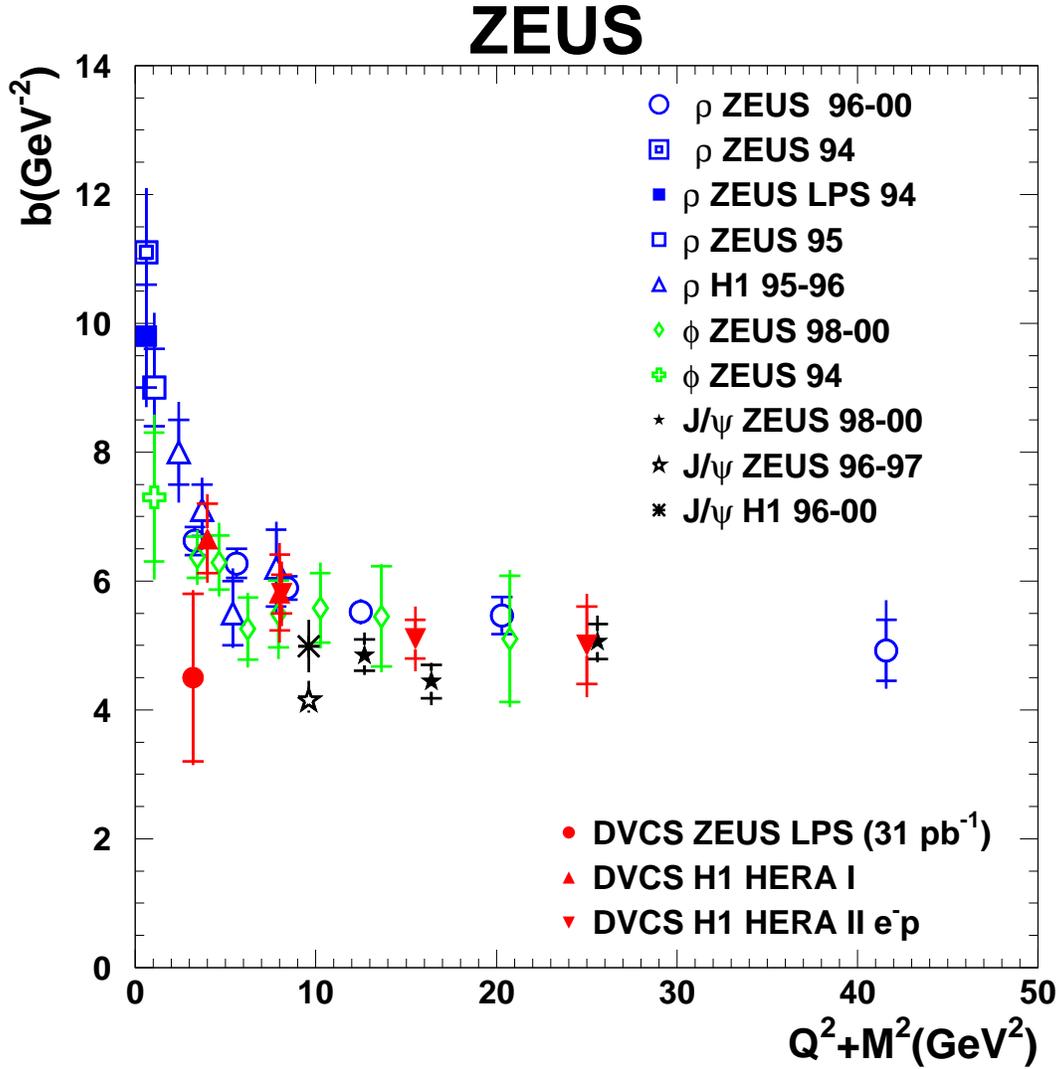}
\end{center}
\vspace{-.5cm}
\caption{
A compilation of the values of the slope b as a function of $Q^2+M^2$ for various exclusive 
processes including the present DVCS measurement. 
The inner error bars represent the statistical uncertainty while the outer
error bars the statistical and systematic uncertainties added in quadrature. 
}
\label{fig:bfignew}

\end{figure}

A compilation of $b$ values as measured for various exclusive
processes~\cite{art:pmc:a2007:1,zfp:c73:253}, 
including the result of this paper, is 
shown  in Fig.~\ref{fig:bfignew} as a function
of $Q^2+M^2$, where $M$ is the mass of the exclusive final state. 
The $b$ value presented here is lower but
consistent with the corresponding vector mesons and H1 DVCS values
at similar scales.
The fast rise of the DVCS cross section with $W$ at $Q^2=2.4$~GeV$^2$ and 
the low value of $b$ at $Q^2=3.2$~GeV$^2$ indicate that the DVCS process 
is a hard process even at low $Q^2$ values.

\section{Summary}
\label{sec-con}

The DVCS cross section has been measured as a function of $Q^2$ and $W$ 
in the region $1.5 < Q^2 < 100$~GeV${}^2$ and $40 < W  < 170$~GeV.
The measured cross section decreases steeply with $Q^2$, showing a
dependence $Q^{-2n}$, with $n = 1.54 \pm 0.05 (\rm stat.)$.  
The $W$ cross section rises with increasing $W$  following a functional form
$W^\delta$, with $\delta = 0.52 \pm 0.09 (\rm stat.)$ and has little 
dependence on  $Q^2$.
  
For the first time, the DVCS differential cross section 
as a function of $t$ was measured by directly tagging the scattered proton.
An exponential behaviour was assumed, the slope parameter 
$b=4.5\pm 1.3 (\rm stat.) \pm 0.4 (\rm syst.)~ \rm{GeV^{-2}}$
was obtained from a fit to the data at $Q^2=3.2$~GeV$^2$ and $W=104$~GeV.
These findings indicate that the DVCS process 
is a hard process even at low $Q^2$.


\section*{Acknowledgements} \label{sec-ack} We thank the DESY Directorate
for their support and encouragement. We are grateful for the support of the
DESY computing and network services. We are specially grateful to the HERA
machine group: collaboration with them was crucial to the successful
installation and operation of the leading proton spectrometer. The design,
construction and installation of the ZEUS detector were made possible
by the ingenuity and effort of many people who are not listed as authors. 

\vfill\eject

\providecommand{\etal}{et al.\xspace}
\providecommand{\coll}{Coll.\xspace}
\catcode`\@=11
\def\@bibitem#1{%
\ifmc@bstsupport
  \mc@iftail{#1}%
    {;\newline\ignorespaces}%
    {\ifmc@first\else.\fi\orig@bibitem{#1}}
  \mc@firstfalse
\else
  \mc@iftail{#1}%
    {\ignorespaces}%
    {\orig@bibitem{#1}}%
\fi}%
\catcode`\@=12

\newpage
\clearpage

\begin{table}[p]
\begin{center}
\begin{tabular}{|c|c|c|}
\hline
\multicolumn{3}{|c|}{$\sigma ^{\gamma^*p\rightarrow \gamma p} $}\\
\hline
$Q^2$ range ($\rm{GeV^2}$) & $Q^2$ ($\rm{GeV^2}$) & $\sigma ^{\gamma^*p\rightarrow \gamma p}$ (nb)\\
\hline
1.5 - 5   & ~~3.25  & 21.28$\pm$0.92$_{-1.34}^{+1.02}$  \\
  ~~~~5 - 10  & ~7.5  & 5.87 $\pm$0.42$_{-0.30}^{+0.14}$   \\
 ~~~10 - 15  & 12.5 & 3.27 $\pm$0.33$_{-0.16}^{+0.07}$\\
 ~~~15 - 25  & 20.0   & 1.23 $\pm$0.21$_{-0.08}^{+0.05}$  \\
 ~~~25 - 40  & 32.5 & 0.55 $\pm$0.18$_{-0.04}^{+0.04}$\\
 ~~~~40 - 100 & 70.0   & 0.16 $\pm$0.07$_{-0.02}^{+0.02}$\\
\hline
\end{tabular}
\caption{
The DVCS cross section, $\sigma^{\gamma ^*p\rightarrow \gamma p}$, as a function of $Q^2$. 
Values are quoted at the centre of each $Q^2$ bin and at $W=104 \gev$. 
The first uncertainty is statistical and the second systematic. 
}
\label{tab:tab1}
\end{center}
\end{table}


\begin{table}[p]
\begin{center}
\begin{tabular}{|c|c|c|}
\hline
\multicolumn{3}{|c|}{$\sigma ^{\gamma^*p\rightarrow \gamma p} $}\\
\hline
$W$ range ($\rm{GeV}$) & $W$ ($\rm{GeV}$) &
 $\sigma ^{\gamma^*p\rightarrow \gamma p} $ (nb)\\
\hline
40 - 60   & 50  &14.47$\pm$1.05$_{-0.88}^{+0.50}$\\
60 - 80   & 70  &20.38$\pm$1.57$_{-1.99}^{+1.01}$\\
80 - 100  & 90  &17.95$\pm$1.35$_{-0.93}^{+0.64}$\\
100 - 120 & 110 &20.65$\pm$1.26$_{-1.16}^{+0.59}$\\
120 - 140 & 130 &26.42$\pm$1.84$_{-0.88}^{+0.79}$\\
140 - 170 & 155 &27.60$\pm$3.74$_{-3.34}^{+2.01}$\\
\hline
\end{tabular}
\caption{
The DVCS cross section, $\sigma^{\gamma ^*p\rightarrow \gamma p}$, as a function of $W$.
Values are quoted at the centre of each $W$ bin and for $Q^2=3.2 \gev^2$. 
The first uncertainty is statistical and the second systematic.  
}
\label{tab:tab2}
\end{center}
\end{table}
\begin{table}[p]
\begin{center}
\begin{tabular}{|c|c|c|c|c|c|}
\hline
\multicolumn{6}{|c|}{$\sigma ^{\gamma^*p\rightarrow \gamma p} $}\\
\hline
$W$ range & $W$  
&$\sigma ^{\gamma^*p\rightarrow \gamma p} $ (nb)
&$\sigma ^{\gamma^*p\rightarrow \gamma p} $ (nb) 
&$\sigma ^{\gamma^*p\rightarrow \gamma p} $ (nb) 
&$\sigma ^{\gamma^*p\rightarrow \gamma p} $ (nb)\\
($\rm{GeV}$)& ($\rm{GeV}$)& $Q^2=2.4\;\Gev^2$& $Q^2=6.2\;\Gev^2$&$Q^2=9.9\;\Gev^2$ &$Q^2=18.0\;\Gev^2$\\
\hline
40 - 65  & 52.5  & 27.06$\pm$3.44$^{+4.37}_{-4.42}$ & & & \\
65 - 90  & 77.5  & 22.36$\pm$3.11$^{+3.40}_{-1.73}$ & & &\\
90 - 115 & 102.5 & 26.49$\pm$1.89$^{+0.89}_{-1.44}$ & & & \\
115 - 140 & 127.5& 35.94$\pm$2.63$^{+1.81}_{-1.89}$ & & & \\
140 - 170 & 155  & 35.72$\pm$9.47$^{+3.01}_{-2.94}$ & 16.93$\pm$2.43$^{+1.37}_{-1.40}$ & 6.15$\pm$1.67$^{+0.51}_{-0.51}$ & 2.21$\pm$0.82$^{+0.18}_{-0.18}$ \\
\hline
\end{tabular}
\caption{
The DVCS cross section, $\sigma^{\gamma ^*p\rightarrow \gamma p}$, as a 
function 
of $W$ in four $Q^2$ ranges.
Values are quoted at the centre of each $W$ bin and for the $Q^2$ values 
listed.  
The first uncertainty is statistical and the second systematic.  
The values for higher $Q^2$ and lower $W$, shown in Fig.~\ref{fig:wfit}, 
are taken from a previous publication~\protect\cite{pl:b573:46} and are
not repeated here.
}
\label{tab:tab3}
\end{center}
\end{table}

\begin{table}[p]
\begin{center}
\begin{tabular}{|c|c|c|}
\hline
\multicolumn{3}{|c|}{$d\sigma ^{\gamma^*p\rightarrow \gamma p}/dt $}\\
\hline
$t$ range ($\rm{GeV^2}$) & $t$ ($\rm{GeV^2}$) & $\sigma ^{\gamma^*p\rightarrow \gamma p}/dt$ (nb/GeV$^2$)\\
\hline
0.08 - 0.19 & 0.14 & 34.6$\pm$9.6$\pm$ 2.4 \\
0.19 - 0.31 & 0.25 & 32.7$\pm$9.4$\pm$ 2.3 \\
0.31 - 0.42 & 0.36 & 19.6$\pm$7.5$\pm$ 1.4 \\
0.42 - 0.53 & 0.47 & 5.7 $\pm$4.1$\pm$ 0.4 \\
\hline
\end{tabular}
\caption{
The DVCS differential cross section, 
$d\sigma^{\gamma ^*p\rightarrow \gamma p}/dt$, as a 
function of $|t|$. 
Values are quoted at the centre of each $|t|$ bin and for 
$Q^2=3.2\gev^2$ and $W=104 \gev$. 
The first uncertainty is statistical and the second systematic. 
}
\label{tab:tab4}
\end{center}
\end{table}


\end{document}